\documentstyle[aps,prb,multicol,epsf]{revtex}

\newcommand{\be}{\begin{equation}}
\newcommand{\ee}{\end{equation}}
\newcommand{\bea}{\begin{eqnarray}}
\newcommand{\eea}{\end{eqnarray}}
\newcommand{\ba}{\begin{array}}
\newcommand{\ea}{\end{array}}

\newcommand{\ep}{\epsilon}
\newcommand{\th}{\theta}
\newcommand{\Ga}{\Gamma}
\newcommand{\ka}{\kappa}

\draft
\begin{document}


\title{Exact shape of the lowest Landau level in a spin-$\frac{1}{2}$ system
with uncorrelated disorders}

\author{Ming-Hsien Tu\cite{MHT} and Min-Fong Yang \cite{MFY} }

\address{
$^*$ Department of Physics,\\
National Tsing Hua University,\\
Hsinchu, Taiwan, R.O.C.,\\
and\\
$^\dagger$ Department of General  Programs,\\
Chang Gung College of Medicine and Technology,\\
Kweishan, Taoyuan, Taiwan, R.O.C.
}

\date{\today}

\maketitle

\begin{abstract}
Using the path-integral approach developed by
Br\'ezin {\it et al.}
[ Nucl.\ Phys.\ B {\bf 235}, 24 (1984) ],
we obtain
an analytical expression for the
density of states of a spin-$\frac{1}{2}$  disordered two-dimensional electron gas
in a strong, perpendicular magnetic field.
The density of states of this system illustrates the
interplay between the Zeeman
splitting of Landau levels and the disorder-induced broadening.
We find that the broadening  and the band splitting of the Landau bands are
enhanced due to the level repulsion from  the mixing of two spin orientations
caused by the random scatterings.
The comparison between the spin-$\frac{1}{2}$ model and the
double-layer system is also discussed.
\end{abstract}



\begin{multicols}{2}

\section{Introduction}

There has been remarkable interest in the problem of a disordered
two-dimensional electron gas in a strong, perpendicular magnetic field
ever since the discovery of the quantum Hall effect \cite{qhe} in such
systems.  Although there have been many studies on this system
\cite{A,C,D,E,Wegner,BGI,H,I,J,K,L,KMT,M,N,O,P,HSW,DKKLee}, however,
only a few rigorous results of the density of states (DOS) for the
disordered electrons are obtained.
The difficulty for obtaining the rigorous expression of DOS
comes from the fact that, in the absence
of the disorder, the energy spectrum is discrete.
Therefore the self-energy of an
electron is real in any finite order of the perturbation theory.
Hence, to obtain the expected disorder-broadened Landau
levels ( LL ), one needs to sum over the entire diagram expansion.  In
the strong-field limit, where only the lowest LL is considered, the
exact DOS of the {\it spinless} electrons was first found by Wegner
\cite{Wegner} for the case of a Gaussian white noise distribution of
the random potential.  Using a functional-integral approach,
Br\'{e}zin, Gross and Itzykson \cite{BGI} were able to re-derive
Wegner's result and, furthermore, to generalize it to the case in
which a non-Gaussian, but still zero-range, distribution of impurities
is assumed.  The crucial point for these exact derivations is the
intrinsic supersymmetry of the problem, by which a mapping from the
original two-dimensional system onto a {\it zero}-dimensional model
becomes possible.  Thus, the summation of the entire diagram expansion
can be achieved.

However, in the real world, electrons carry spins. As a result, the
effects of mixing between LL's with different spin quantum numbers
would become important, when the Zeeman splitting is smaller than the
width of each disorder-broadened LL.  Hence, the generalization of
above results to the spin-$\frac{1}{2}$ case is of pratical interest.
However, as was pointed out in the last section of Ref.\cite{BGI}, the
identity ensuring the existence of the supersymmetry fails to hold in
the two-component case and implying that a straightward generalization
to the spin-$\frac{1}{2}$ case appears impossible in general.
Nevertheless, we find that an exact expression of DOS can be reached
provided that the spin-$\frac{1}{2}$ electrons are influenced by
the random potential scattering {\it and the random spin-flip
scattering}, which are both assumed to be the zero-range white-noise
distribution.  We find that the energy difference of the splitted
bands is {\it larger} than the Zeeman splitting energy.  Moreover,
when the Zeeman splitting is turned off, the width of the
disorder-broadened band is {\it larger} than, rather than identical
to, that of the spinless case.  Thus an interesting interplay between
the Zeeman splitting of Landau levels and the disorder-induced
broadening can be demonstrated in our spin-$\frac{1}{2}$ system.
In Section II, we introduce the model Hamiltonian and outline the
functional-integral formulation. In Sec.III, the exact results of DOS
are presented, which show the enhancement of the splitting and the
band width. In Sec.IV, we discuss the asymptotic behavior of DOS when
Zeeman splitting is large. Section V is devoted to the
conclusion and the comparison between the spin-$\frac{1}{2}$ model
and the double-layer system.

\section{The Model and the Functional-Integral Formulation}

We consider the following Hamiltonian,
\begin{eqnarray}
H   &=&H_0 + U_1({\bf r}) + U_2({\bf r}) \tau_1  ,
\label{ham} \\
H_0 &=&\frac{1}{2m}({\bf P\/} - e {\bf A\/})^{2}
        - \frac{1}{2} g\ \mu_{\rm B} B \tau_3  ,
\end{eqnarray}
where ${\bf A}({\bf r})=(-By/2,Bx/2)$ is the vector potential of the
constant magnetic field, ${\bf B}({\bf r}) = B \hat{\bf z}$, in the
symmetric gauge, $g$ is the gyromagnetic ratio, $\mu_{\rm B}$ is the
Bohr magneton, $\tau_1$ and $\tau_3$ are Pauli matrices.  The random
potential scattering and the random spin-flip scattering are denoted
by $U_1({\bf r\/})$ and $U_2({\bf r\/})\tau_1$, respectively.  The
latter term can be visualized as a simplified description of the
scattering by magnetic impurities, or the spin-orbit scattering.
\cite{HAMG} ( A similar model has been considered by Wang {\it et al.}
\cite{WLW} to study the effects of Landau level mixing ).

We first outline briefly the procedure of
calculating DOS developed by Br\'ezin {\it et al.} \cite{BGI}.
Assuming that the magnetic field is so strong that the
random scatterings cannot
induce transitions between different Landau levels,  the
averaged DOS of the lowest LL can be written as \cite{Wegner}
\begin{equation}
\rho(E)=-\frac{1}{\pi}\mbox{Im}
         \overline{\hbox{Tr}
          \left\{ \langle {\bf r}|\frac{1}{E-H+i0}|{\bf r}\rangle \right\}  },
\label{dos}
\end{equation}
where Tr denotes the trace operation over the spin indices and the bar
indicates averaging over all configurations of the random potentials.
Using the functional-integral approach, \cite{BGI} the matrix element
of the resolvent after random average becomes
\bea
&&\overline{\hbox{Tr}
         \left\{ \langle {\bf r}|\frac{1}{E-H+i0}|{\bf r}\rangle \right\}  }
=-i e^{-\frac{1}{2}\kappa^2|z|^2}  \nonumber \\
&&\times \int \prod_{\sigma=\pm}
  Du_{\sigma} Du^*_{\sigma} Dv_{\sigma} D\bar{v}_{\sigma}
  \sum_{\sigma=\pm}
  \left( u^*_{\sigma} u_{\sigma}
         + \bar{v}_{\sigma} v_{\sigma} \right)   e^S,
\label{reso}
\eea
with the ``action" $S$ given by
\begin{eqnarray}
S&=&i\int dzdz^* e^{-\frac{1}{2}\kappa^2|z|^2}
    \sum_{\sigma=\pm} (\ep-\sigma \bar{g})
  \left( u^*_{\sigma} u_{\sigma}
         + \bar{v}_{\sigma} v_{\sigma} \right)
                                       \nonumber \\
  && + \int dzdz^* f_1\left[ e^{-\frac{1}{2}\kappa^2|z|^2}
    \sum_{\sigma=\pm}
  \left( u^*_{\sigma} u_{\sigma}
         + \bar{v}_{\sigma} v_{\sigma} \right) \right] \\
 && + \int dzdz^*f_2\left[ e^{-\frac{1}{2}\kappa^2|z|^2}
    \sum_{\sigma=\pm}
  \left( u^*_{\sigma} u_{-\sigma}
  + \bar{v}_{\sigma} v_{-\sigma} \right) \right],
                                       \nonumber
\end{eqnarray}
where $\kappa^2 = eB/\hbar$,
$\epsilon = E - \frac{1}{2} \hbar \frac{eB}{m}$,
$\bar{g} = \frac{1}{2} g\ \mu_{\rm B} B$, and $z=x+iy$
denotes the complex coordinates on the plane.  The holomorphic bosonic
( fermionic ) fields and their adjoints are denoted by $u_{\pm}(z)$ (
$v_{\pm}(z)$ ) and $u^*_{\pm}(z^*)$ ( $\bar{v}_{\pm}(z^*)$ )
respectively.  The Fourier transform of the distribution function
$P_j(U_j)$ at a single site, $f_j(\alpha)$, is defined by
\begin{equation}
f_j(\alpha)=\ln\int dU_j P_j(U_j) e^{-i\alpha U_j},\qquad f_j(0)=0,
             \qquad j=1,2.
\end{equation}
In this paper, we  restrict ourselves to the case where
both $P_1$ and $P_2$ are white-noise distributions of the same
strength, $w$, i.e.,
\be
P_1(U_1) = \exp(-\frac{|U_1|^2}{2w}), \qquad
P_2(U_2) = \exp(-\frac{|U_2|^2}{2w}),
\ee
such that $f_1(\alpha) = f_2(\alpha) = -\frac{w}{2} \alpha^2$.
To examine whether the supersymmetry exists in this path-integral
representation, we introduce the holomorphic superfields and their
adjoints for each spin orientation,
\bea
\Phi_{\pm} (z,\theta)
&=& u_{\pm}(z) + \frac{1}{\sqrt{2}}\kappa\theta v_{\pm}(z) , \\
\bar{\Phi}_{\pm} (z^*,\bar{\theta})
&=& u^*_{\pm}(z^*) + \frac{1}{\sqrt{2}}\kappa \bar{v}_{\pm}(z^*) \bar{\theta} ,
\eea
where $\th$ and $\bar{\th}$ are the Grassmann coordinates satisfying
\be
\int d\th d\bar{\th} = \int d\th d\bar{\th}  \th =
\int d\th d\bar{\th} \bar{\th} = 0, \qquad
\int d\th d\bar{\th} \bar{\th}\th = \frac{1}{\pi}.
\ee
\end{multicols}
Consequently, the `` action " $S$ can be expressed in terms of the
superfields
$\Phi_{\pm}(z,\theta)$ and $\bar{\Phi}_{\pm}(z^*,\bar{\theta})$ as
\bea
S&=&\frac{2\pi i}{\kappa^2}\int dzdz^*d\theta d\bar{\theta}
    e^{-\frac{1}{2}\kappa^2(|z|^2+\th\bar{\th})}
\left[ \ep (\bar{\Phi}_+ \Phi_+ + \bar{\Phi}_- \Phi_- )
     -\bar{g} (\bar{\Phi}_+ \Phi_+ - \bar{\Phi}_- \Phi_- ) \right] \nonumber \\
 && -\frac{w}{2} \int dzdz^*
\left[  e^{-\frac{1}{2}\kappa^2|z|^2}\frac{2\pi}{\kappa^2}
    \int d\theta d\bar{\theta} e^{-\frac{1}{2}\kappa^2\th\bar{\th}}
      (\bar{\Phi}_+ \Phi_+ + \bar{\Phi}_- \Phi_- ) \right]^2
\label{S}\\
 && -\frac{w}{2} \int dzdz^*
\left[  e^{-\frac{1}{2}\kappa^2|z|^2}\frac{2\pi}{\kappa^2}
    \int d\theta d\bar{\theta} e^{-\frac{1}{2}\kappa^2\th\bar{\th}}
      (\bar{\Phi}_+ \Phi_- + \bar{\Phi}_- \Phi_+ ) \right]^2 . \nonumber
\eea
\begin{multicols}{2}
Obviously, the first integral in Eq.(\ref{S}) is invariant under
superspace rotations,
\bea
&&\delta z=\bar{\omega}\th,\qquad \delta z^*=\bar{\th} \omega ,\\
&&\delta\th=\omega z,\qquad \delta\bar{\th}=z^*\bar{\omega} ,
\eea
which preserves the quadratic term $(zz^*+\th\bar{\th})$.  However, as
$z$ and $\th$ are not treated on the same footing in
the second and the third integrals of Eq.(\ref{S}),
the invariance for these integrals is not necessarily
promised.  For the spinless electrons described by the superfield
$\Phi$, the following identity
\be
\left[\frac{2\pi}{\kappa^2} \int d\th d\bar{\th}
    e^{-\frac{1}{2}\kappa^2\th\bar{\th}} \bar{\Phi} \Phi  \right]^n
 = \frac{1}{n} \frac{2\pi}{\kappa^2} \int d\th d\bar{\th}
    e^{-\frac{1}{2}n\kappa^2\th\bar{\th}} ( \bar{\Phi} \Phi )^n
\label{id}
\ee
ensures the supersymmetry and then the exact DOS of spinless electrons.
\cite{BGI} However, as was pointed out in Ref.\cite{BGI}, the
generalization of
Eq.(\ref{id}) to the two-component case, say,
$\Phi \to (\bar{\Phi}_+ \Phi_+ + \bar{\Phi}_- \Phi_- )$,
is not allowed when $n >1$ as the
left-hand side of Eq.(\ref{id}) contains the four fermion
interactions, $\bar{v}_+ v_+ \bar{v}_- v_-$, whereas the right-hand
side of Eq.(\ref{id}) does not.  Hence, the supersymmetry, which
enables one to obtain an exact expression of DOS, does not seem to
exist in the spin-$\frac{1}{2}$ case.  However, when the two random
distributions $P_1$ and $P_2$ are both Gaussian and of the same
strength, these four fermion interactions cancel each other in the sum
of the second and third integrals in Eq.(\ref{S}).  This cancellation
can be easily illustrated by the following changes of variables for
the superfields:
\bea
\Phi_1&=& \frac{1}{\sqrt{2}} ( \Phi_+ + \Phi_- ), \qquad
\bar{\Phi}_1= \frac{1}{\sqrt{2}} ( \bar{\Phi}_+ + \bar{\Phi}_- ),
\label{change1} \\
\Phi_2&=& \frac{1}{\sqrt{2}} ( \Phi_+ - \Phi_- ), \qquad
\bar{\Phi}_2= \frac{1}{\sqrt{2}} ( \bar{\Phi}_+ - \bar{\Phi}_- ),
\label{change2}
\eea
therefore, a supersymmetric `` action " is reached,
\bea
S&=&\frac{2\pi i}{\kappa^2}\int dzdz^*d\theta d\bar{\theta}
     e^{-\frac{1}{2}\kappa^2(|z|^2+\th\bar{\th})}
                                        \nonumber \\
  &&\times \left[ \ep (\bar{\Phi}_1 \Phi_1 + \bar{\Phi}_2 \Phi_2 )
     -\bar{g} (\bar{\Phi}_1 \Phi_2 + \bar{\Phi}_2 \Phi_1 ) \right]
                                        \nonumber \\
&&-\frac{w\pi}{\kappa^2}\int dzdz^*d\th d\bar{\th}
     e^{-\kappa^2(|z|^2+\th\bar{\th})}
\label{superS} \\
&& \times \left[ (\bar{\Phi}_1 \Phi_1)^2
          + (\bar{\Phi}_2 \Phi_2 )^2 \right] .   \nonumber
\eea
In the derivation of Eq.(\ref{superS}),
the identity Eq.(\ref{id}) is used for {\it each} $\Phi_1$ and
$\Phi_2$ in the intermediate step, {\it in which no four fermion term
is involved}.  Because of the dimensional reduction, $d \to d-2$, due
to the existence of the supersymmetry in Eq.(\ref{superS}), the
calculations of the averaged resolvent, Eq.(\ref{reso}), can be
reduced to evaluate the following zero-dimensional integral,
\cite{BGI}
\bea
&&\overline{\hbox{Tr} \left\{ \langle {\bf r}|\frac{1}{E-H+i0}|{\bf r}\rangle
                             \right\} } \nonumber \\
&=&-i \frac{\int d\phi_1 d\phi^*_1 d\phi_2 d\phi^*_2
                   ( \phi^*_1 \phi_1 + \phi^*_2 \phi_2 ) e^{S_0} }
          {\int d\phi_1 d\phi^*_1 d\phi_2 d\phi^*_2 e^{S_0}     },
\label{0dint}
\eea
where $S_0$ is defined by
\bea
S_0
&=&i \frac{2\pi}{\kappa^2} \ep ( \phi^*_1 \phi_1 + \phi^*_2 \phi_2 )
  -i \frac{2\pi}{\kappa^2} \bar{g} ( \phi^*_1 \phi_2 + \phi^*_2 \phi_1 )
                        \nonumber \\
&&-\frac{w\pi}{\kappa^2}
  \left[ ( \phi^*_1 \phi_1)^2 + (\phi^*_2 \phi_2 )^2 \right].
\label{S0}
\eea
Combining Eqs.(\ref{dos}), (\ref{0dint}) and (\ref{S0}), we have
the following exact expression for DOS
\be
\rho(E)=\frac{\kappa^2}{2\pi^2}\mbox{Im}\frac{\partial}{\partial \ep}
\ln Z_0
\label{exact}
\ee
with
\be
Z_0 = \int d\phi_1 d\phi^*_1 d\phi_2 d\phi^*_2 e^{S_0}.
\ee
It is convenient to rewrite $Z_0$ in a different form.  We first
rescale $\phi_i$ by
$\phi^{\prime}_i = \sqrt{\frac{2\pi}{\kappa^2}} \phi_i$,
and subsequently decouple the quartic terms in $S_0$ with the
help of gaussian integral over a pair of auxiliary variables,
$\lambda_1$ and $\lambda_2$.  Finally we calculate the remaining
integral over $\phi^{\prime}_i$ to obtain
\begin{eqnarray}
Z_0
&=&\frac{(i\pi)^2}{\pi (\frac{2\pi}{\kappa^2})^2 \Gamma^2}
    \int_{-\infty}^{\infty}\int_{-\infty}^{\infty}
        d\lambda_1 d\lambda_2
        \frac{e^{-(\lambda_1^2 + \lambda_2^2 )/\Gamma^2}}
             {(\epsilon+\lambda_1)(\epsilon+\lambda_2)-\bar{g}^2}
             \nonumber \\
&=&Z_0^{\rm R} + i Z_0^{\rm I}
\label{Z0}
\end{eqnarray}
with the real and the imaginary parts given by
\begin{eqnarray}
Z_0^{\rm R}
&=& - \frac{2}{\sqrt{\pi}} \frac{\kappa^4}{2\Gamma^2}
 \int_0^{\infty} dy \frac{e^{-y^2}}{\sqrt{y^2 + 2\bar{g}^2/\Gamma^2}}
                                        \nonumber \\
&&\times \left( e^{-\nu_-^2} \int_0^{\nu_-} dx e^{x^2}
         -e^{-\nu_+^2} \int_0^{\nu_+} dx e^{x^2}   \right) ,
\label{zr} \\
Z_0^{\rm I}
&=&\frac{\kappa^4}{2\Gamma^2}
  \int_0^{\infty} dy \frac{e^{-y^2}}{\sqrt{y^2 + 2\bar{g}^2/\Gamma^2}}
   \left( e^{-\nu_-^2} - e^{-\nu_+^2} \right) ,
\label{zi}
\end{eqnarray}
where the magnetic-field-dependent width
$\Gamma = \sqrt{w\kappa^2/\pi}$,
$\nu_\pm = \sqrt{2}\epsilon / \Gamma
\pm \sqrt{y^2 + 2\bar{g}^2/\Gamma^2}$,
and the changes of variables,
$x = ( \lambda_1 + \lambda_2 )/\sqrt{2}$ and
$y = ( \lambda_1 - \lambda_2 )/\sqrt{2}$,
are introduced.

\section{Results of Various Density of States }

The general behavior of DOS is shown in Fig.1.  Only the part for
$\epsilon \ge 0$ is shown, since $\rho(E)$ is symmetric with respect
to the band center ( $\epsilon=0$ ).  \cite{note} Figure 1(a) shows,
as expected, that the splitting of the disordered Landau bands
increases as $g$ increases.  To get more insight from our result, we
depict in Fig.1(b) by shifting horizontally each curve of Fig.1(a)
with the Zeeman energy, $\bar{g} = \frac{1}{2} g\ \mu_{\rm B} B$.  It
becomes apparent that, as $g$ increases, the shifted curves evolve
to a single curve having the simialar form of DOS in the spinless
case, $\rho_0 (E)$.  \cite{BGI} However, two unexpected results
emerge:  first, the peak values of the splitted bands occur at a {\it
larger} $\epsilon$, rather than exactly at $\epsilon = \bar{g}$ as one
expects naively; secondly, if there is no Zeeman splitting ( i.e.,
$g=0$ ), from Eqs.(\ref{Z0})-(\ref{zi}) and (\ref{exact}), we obtain
\begin{eqnarray}
\rho(E)|_{g=0}
=\sqrt{2} \frac{\kappa^2}{2\pi^2} \frac{2}{\Gamma_0 \sqrt{\pi}}
   \frac{e^{\epsilon^2/2\Gamma_0^2}}
           {1
            + \left( \frac{2}{\sqrt \pi} \int_0^{\epsilon/\sqrt{2}\Gamma_0} dx
                     e^{x^2} \right)^2 }.
\label{g0}
\end{eqnarray}
where $\Gamma_0 = \sqrt{w\kappa^2/2\pi} = \Gamma/ \sqrt{2}$
is the corresponding magnetic-field-dependent width in the spinless case.
( See Eqs.(46a) and (46b) in Ref.\cite{BGI}. )
Thus the DOS in the spin-$\frac{1}{2}$ system without Zeeman splitting is {\it not}
simply twice as large as that in the spinless case.
In our case, the peak value at the
band center
( $\epsilon = 0$ ) of Eq.(\ref{g0}) is merely $\sqrt 2$
times of that in the spinless case, and the width of $\rho(E)$ is
$\sqrt{2}$ times wider. It is because
the random scatterings mix the two spin degrees of
freedom, as indicated by the quartic terms
in Eq.(\ref{S}).
Therefore, we may expect a
level splitting
due to the usual repulsion of eigenvalues by these terms, \cite{HAMG} and thus
the broadening
of the Landau bands is enhanced when $g=0$, while the band splitting becomes
larger for $g \neq 0$.

\begin{figure}[hbt]
\centerline{\epsfysize=4cm \epsfxsize=7.5cm
\epsfbox{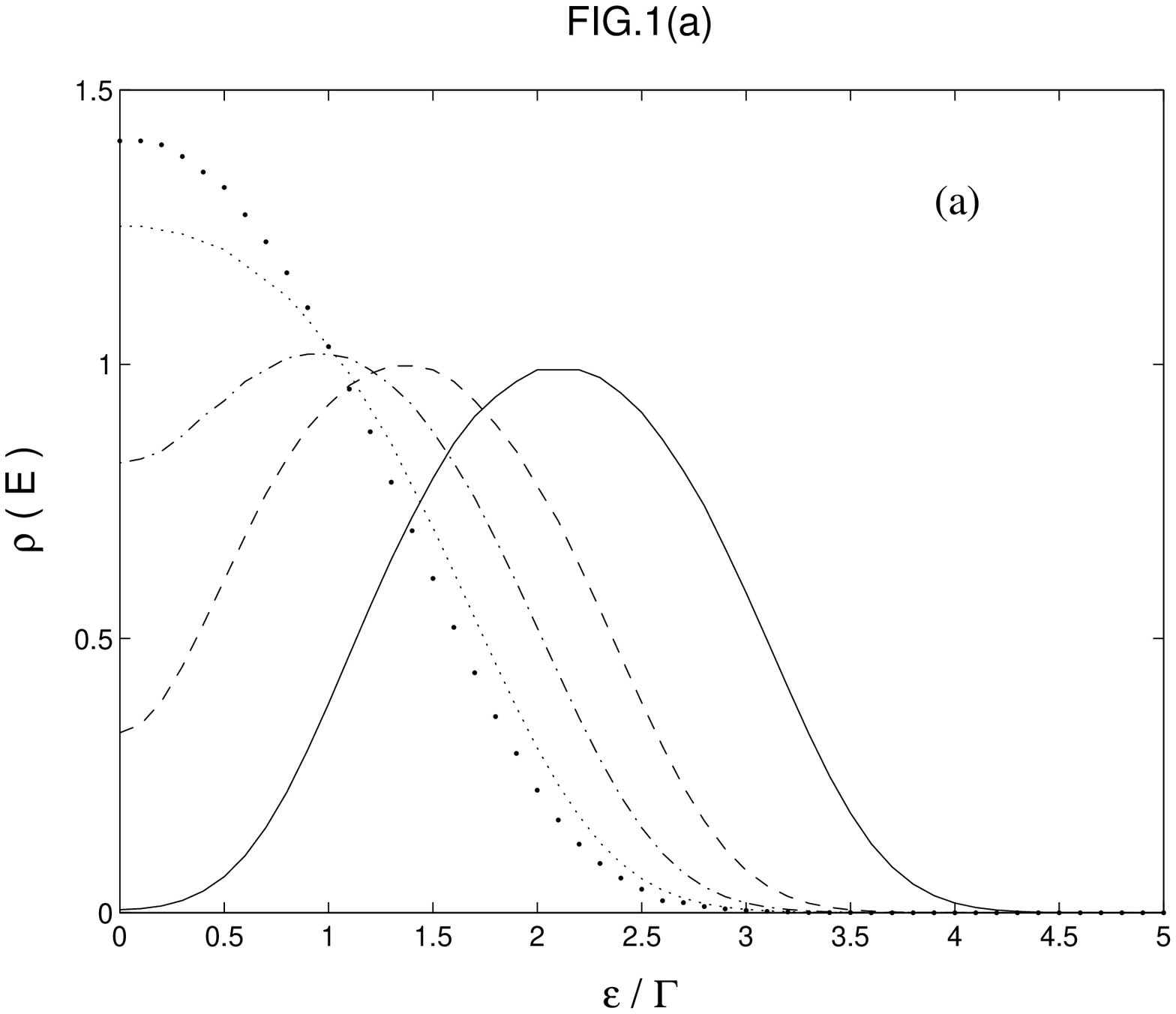}}
\centerline{\epsfysize=4cm \epsfxsize=7.5cm
\epsfbox{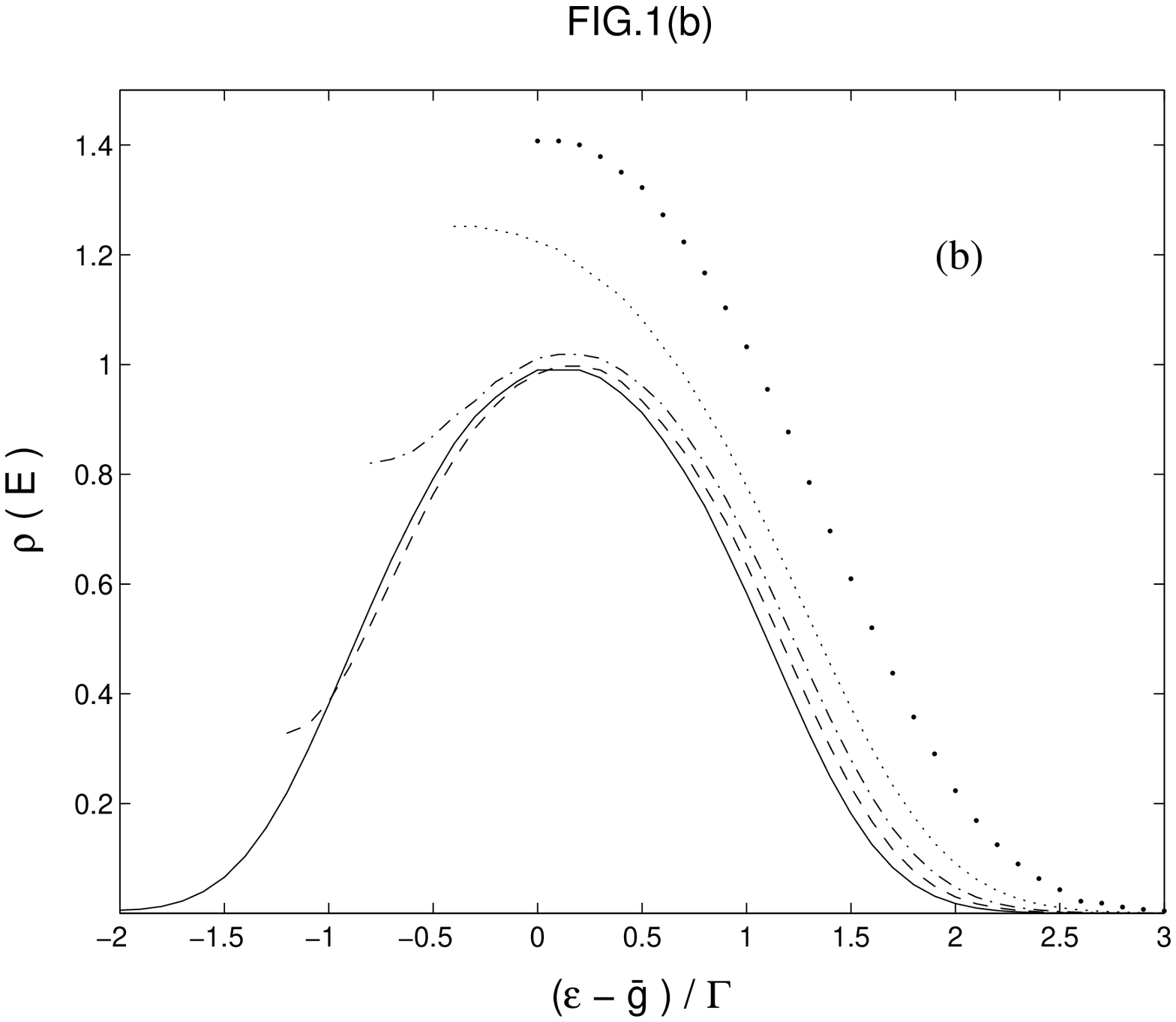}}
{Fig.1 :
(a)
The averaged DOS for a spin-$\frac{1}{2}$ system in units of the peak
value of the spinless case,
$\rho_0 (E) |_{\epsilon = 0} =
\protect\sqrt{2} \ \kappa^2/\pi^{5/2}\Gamma$,
for values of $\bar{g}/\Gamma=$0.0 ( points ), 0.4 ( dotted line ),
0.8 ( dashdotted line ), 1.2 ( dashed line ), and 2.0 ( soild line ).
(b)
The curves in (a) are shifted horizontally by the Zeeman
splitting energy, $\bar{g} = \frac{1}{2} g\ \mu_{\rm B} B$.
}
\end{figure}

Moreover, the exact expression of the disorder-averaged DOS for each spin
component, which is denoted by $\rho_\uparrow (E)$
( $\rho_\downarrow (E)$  ) for up ( down ) spin, can be obtained for the
present model. Parallel to the discussion for the derivation of the sum of
DOS's of two spin components, we have
\be
\rho (E) = \rho_\uparrow (E) + \rho_\downarrow (E)  .
\ee
On the other hand,
their difference can be expressed by ( c.f. Eq.(\ref{exact}) )
\bea
\bar{\rho} (E)
&=& \rho_\uparrow (E) - \rho_\downarrow (E)  \nonumber \\
&=&- \frac{\kappa^2}{2\pi^2}\mbox{Im}
     \frac{\partial}{\partial \bar{g}} \ln Z_0.
\label{exact2}
\eea
Consequently, we have
\bea
\rho_\uparrow (E)
&=& \frac{1}{2} ( \rho (E) + \bar{\rho} (E) ),
\label{defup}\\
\rho_\downarrow (E)
&=& \frac{1}{2} ( \rho (E) - \bar{\rho} (E) ).
\label{defdn}
\eea
By a similar reasoning of the mirror symmetry about the band center
( $\epsilon = 0$ ) for $\rho (E)$, \cite{note}
one can easily show that
\be
\bar{\rho} (E)
\stackrel{\epsilon \to -\epsilon}{\longrightarrow} - \bar{\rho} (E),
\label{mirror}
\ee
and then we have
\bea
\rho_\uparrow (E)
&\stackrel{\epsilon \to -\epsilon}{\longrightarrow}& \rho_\downarrow (E), \\
\rho_\downarrow (E)
&\stackrel{\epsilon \to -\epsilon}{\longrightarrow}& \rho_\uparrow (E).
\eea
With these relations in mind, we show our results of
$\rho_\uparrow (E)$ and $\rho_\downarrow (E)$ only for
$\epsilon \ge 0$ in Figs.2(a)
and (b).  We find that, as $g$ increases, $\rho_\downarrow (E)$
diminishes, while $\rho_\uparrow (E)$ evolves to a similar form of
$\rho_0 (E)$.  The latter can be seen more clearly, if we shift
horizontally each curve in Fig.2(a) with $\bar{g}$ as shown in Fig.3.
The interesting points are that :  first, the peak value of
$\rho_\uparrow (E)$ ( $\rho_\downarrow (E)$ ) does {\it not} occur at
$\epsilon = \bar{g}$ ( $\epsilon = - \bar{g}$ ); secondly,
$\rho_\uparrow (E)$ and $\rho_\downarrow (E)$ at the band center
diminish very fast so that a small bump of $\rho_\downarrow (E)$ (
$\rho_\uparrow (E)$ ) is developed around $\epsilon \sim \bar{g}$ (
$\epsilon \sim - \bar{g}$ ).  Again, these results can be explained by
the level repulsion from the mixing of two spin degrees of freedom.

\begin{figure}[hbt]
\centerline{\epsfysize=4cm \epsfxsize=7.5cm
\epsfbox{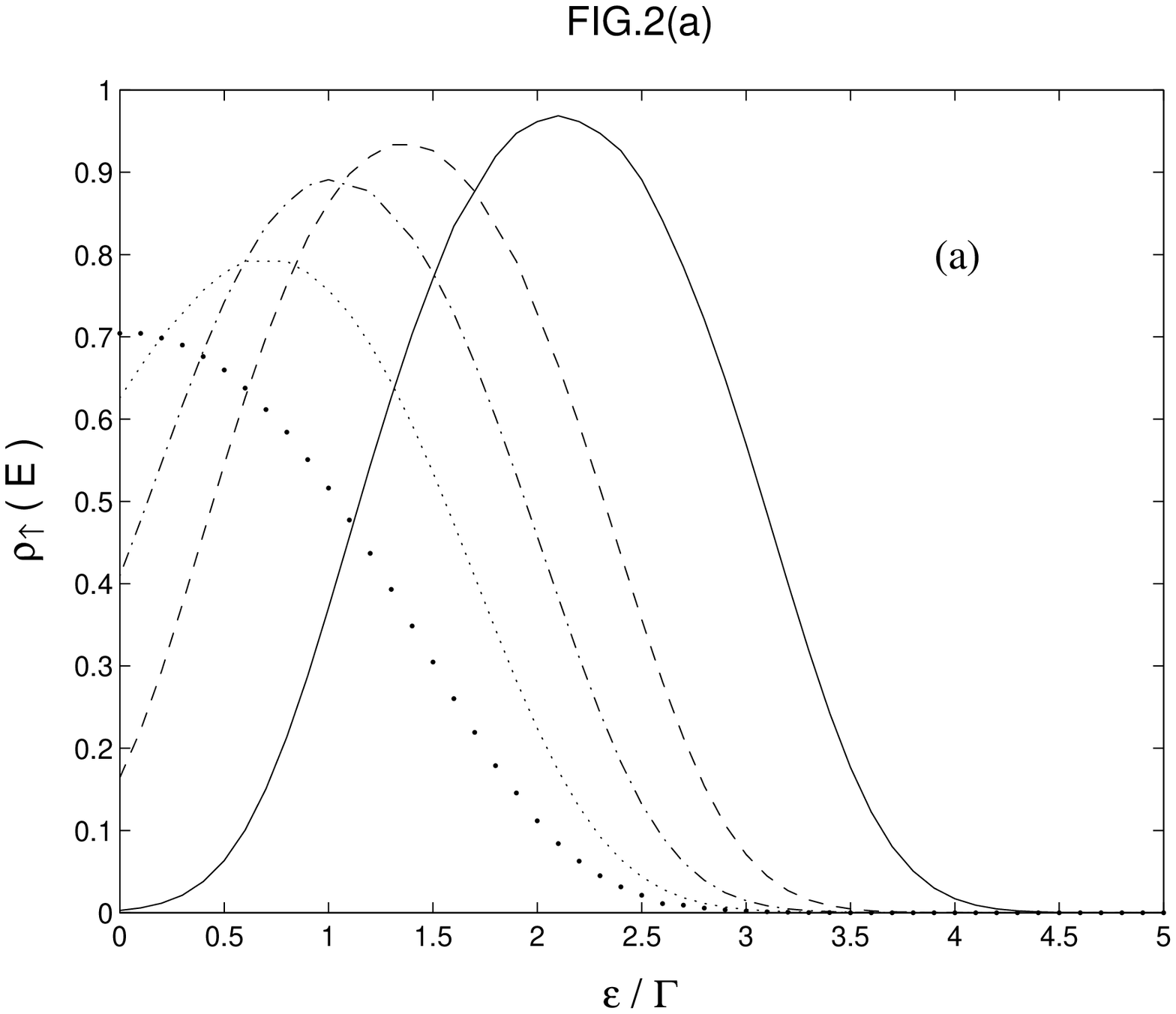}}
\centerline{\epsfysize=4cm \epsfxsize=7.5cm
\epsfbox{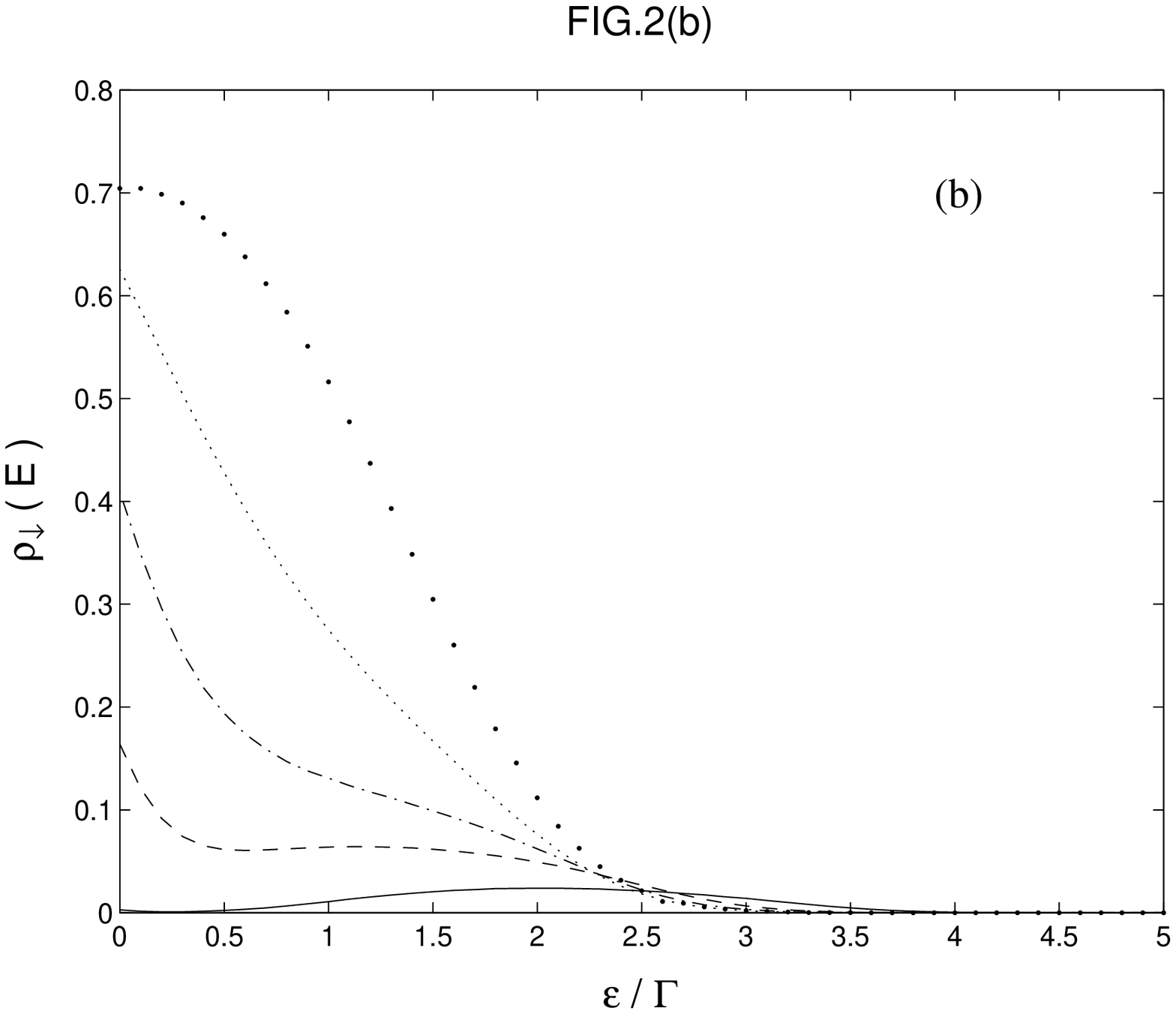}}
{Fig.2 :
The averaged DOS for (a) the spin-up component, $\rho_\uparrow (E)$,
and (b) the spin-down component, $\rho_\downarrow (E)$, in units of
the peak value of the spinless case,
$\rho_0 (E) |_{\epsilon = 0} =
\protect\sqrt{2} \ \kappa^2/\pi^{5/2}\Gamma$,
for values of $\bar{g}/\Gamma=$0.0 ( points ), 0.4 ( dotted line ),
0.8 ( dashdotted line ), 1.2 ( dashed line ), and 2.0 ( soild line ).
}
\end{figure}

\begin{figure}[hbt]
\centerline{\epsfysize=7.5cm
\epsfbox{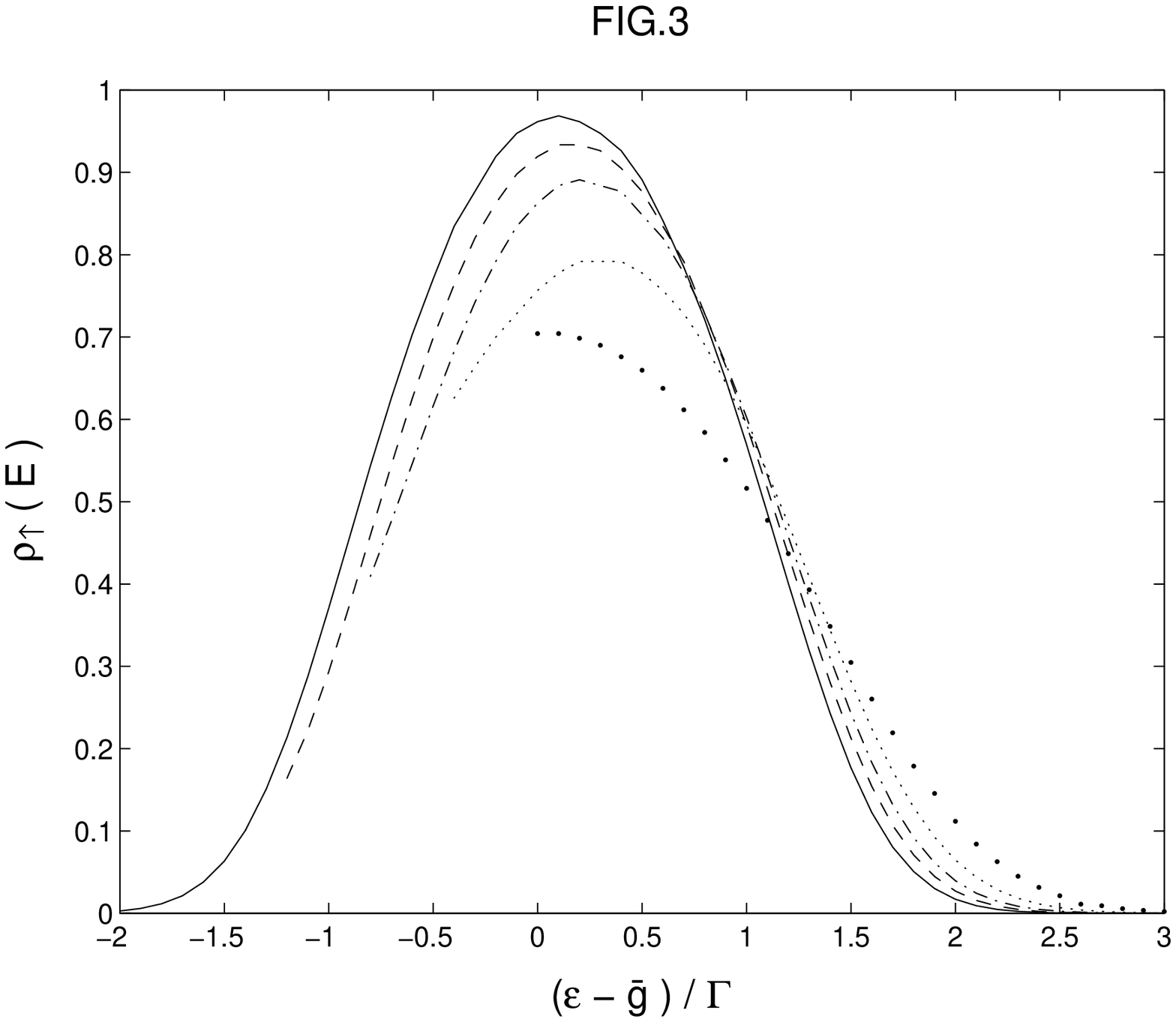}}
{Fig.3 :
The shifted DOS for the spin-up component, $\rho_\uparrow (E)$, in
Fig.2(a) by the Zeeman splitting energy,
$\bar{g} = \frac{1}{2} g\ \mu_{\rm B} B$.
}
\end{figure}

\section{Asymptotic Behavior of Density of States for Large Zeeman
Splitting}

In the following, we consider the asymptotic behavior of DOS in
large-$\bar{g}$ limit ( $\bar{g}/\Gamma \gg 1$ ) and give the
analytical descriptions of the features mentioned above.  Especially,
we will focus our attention on the three regions of energy :  (1) near
the location of the peak; (2) at the band center; (3) toward the band
tails.

Due to the exponential factor, $\hbox{exp}(-y^2)$, in the integrands
of Eqs.(\ref{zr}) and (\ref{zi}), the characteristic value of $y$,
being of order 1, is much smaller than $\bar{g}/\Ga$ when
$\bar{g}/\Gamma \gg 1$.  Thus we can neglect $y^2$ in
$\sqrt{y^2+2\bar{g}^2/\Ga}$, and then
$\nu_{\pm} \cong \sqrt{2}x_{\pm}$, where $x_{\pm}=(\ep\pm\bar{g})/\Ga$
is independent of $y$.  In this case, the integrations over $y$ in
Eqs.(\ref{zr}) and (\ref{zi}) can merely contribute constant factors.
Therefore, the real part and the imaginary part of $Z_0$ become
\bea
Z_0^{\rm R} &\cong&
-\frac{\ka^4}{2\sqrt{2}\Ga\bar{g}} (F_--F_+),
\label{zr1}\\
Z_0^{\rm I} &\cong&
\frac{\ka^4}{2\sqrt{2}\Ga\bar{g}} \frac{\sqrt{\pi}}{2} (f_--f_+),
\label{zi1}
\eea
where
$f_{\pm}=e^{-2x_{\pm}^2}$, and
$F_{\pm}=e^{-2x_{\pm}^2}\int_0^{\sqrt{2}x_{\pm}}dxe^{x^2}.$ From
Eqs.(\ref{exact}) and (\ref{exact2}), we have \bea
\rho(E)&\cong&
-\frac{\sqrt{2}\ka^2}{\pi^{5/2}\Ga}\frac{4\sqrt{2}\frac{\bar{g}}
{\Ga}(f_+F_--f_-F_+)}{\frac{4}{\pi}
(F_--F_+)^2+(f_--f_+)^2},\\
\bar{\rho}(E)&\cong&
\frac{\sqrt{2}\ka^2}{\pi^{5/2}\Ga}
\frac{4\sqrt{2}
[\frac{\ep}{\Ga}(f_+F_--f_-F_+)+\frac{\sqrt{2}}{4}
(f_--f_+)]}
{\frac{4}{\pi}(F_--F_+)^2+(f_--f_+)^2}.
\eea
Consequently, the spin-up and spin-down components of DOS become
\bea
\rho_{\uparrow}(E)&\cong&
\frac{\sqrt{2}\ka^2}{\pi^{5/2}\Ga}
\frac{2\sqrt{2}
[x_-(f_+F_--f_-F_+)+\frac{\sqrt{2}}{4}
(f_--f_+)]}
{\frac{4}{\pi}(F_--F_+)^2+(f_--f_+)^2},
\label{upay}\\
\rho_{\downarrow}(E)&\cong&
-\frac{\sqrt{2}\ka^2}{\pi^{5/2}\Ga}
\frac{2\sqrt{2}
[x_+(f_+F_--f_-F_+)+\frac{\sqrt{2}}{4}
(f_--f_+)]}
{\frac{4}{\pi}(F_--F_+)^2+(f_--f_+)^2}.
\label{downay}
\eea
It can be easily shown that the mirror symmetry about $\epsilon = 0$
still remains in the large-$\bar{g}$ limit.

Let us first consider the behavior of $\rho_{\uparrow}(E)$ near the
location of its peak value.  From Fig.  3, we find that the peak of
$\rho_{\uparrow}(E)$ occurs at $\epsilon = \bar{g}_0$, which is
somewhat larger than $\bar{g}$, and the deviation of $\bar{g}_0$ from
$\bar{g}$ decreases as $\bar{g}$ increases.  Hence,
$\rho_{\uparrow}(E)$ can be approximated to its second-order Taylor
polynomial about $\epsilon = \bar{g}$, and then we have
\be
\rho_{\uparrow}(E)\cong
A ( \frac{\epsilon-\bar{g}}{\Gamma} - B )^2 + C
\label{ABC}
\ee
with
\bea
A&=&\frac{\sqrt{2}\ka^2}{\pi^{5/2}\Ga}2(1-\frac{4}{\pi})+
O(\frac{1}{(\bar{g}/\Ga)^2}),
\label{A} \\
B&=&\frac{1}{8}\frac{1}{(\bar{g}/\Ga)}+O(\frac{1}{(\bar{g}/\Ga)^3}),
\label{B} \\
C&=&\frac{\sqrt{2}\ka^2}{\pi^{5/2}\Ga}(1-\frac{1}{32(\bar{g}/\Ga)^2})+
O(\frac{1}{(\bar{g}/\Ga)^4}),
\label{C}
\eea
where the asymptotic expansion
\be
\int_0^{z}dxe^{x^2}\cong
\frac{e^{z^2}}{2z}(1+\frac{1}{2z^2})
\label{asympt}
\ee
for large $z$ is used for $F_{\pm}$ in Eq.(\ref{upay}). Thus
we find that the peak of $\rho_\uparrow (E)$ occurs at
$\bar{g}_0 / \Ga = \bar{g} / \Ga + B
\cong\ \bar{g} / \Ga + \frac{1}{8(\bar{g}/\Ga)}$
with the peak value
$\rho_\uparrow (E) |_{\epsilon = \bar{g}_0} = C
\cong \frac{\sqrt{2}\ka^2}{\pi^{5/2}\Ga}
      (1-\frac{1}{32(\bar{g}/\Ga)^2}).$
For $\bar{g}/\Ga=2.0$, we have
$\bar{g}_0/\Ga\cong 2.06$ and $C\cong 0.98\frac{\sqrt{2}\ka^2}
{\pi^{5/2}\Ga}$, which agrees with the solid line in Fig.2(a).
Moreover, in the limit $\bar{g}\to\infty$, we have
$\bar{g}_0\to\bar{g}$, and $\rho_{\uparrow}(E)$ around
$\epsilon = \bar{g}_0$ converges to
\be
\rho_{\uparrow}(E)\cong
 \frac{\ka^2}{\pi^{5/2}\Ga_0}
\left [ 1 + (1-\frac{4}{\pi}) (\frac{\ep^{\prime}}{\Ga_0})^2) \right ],
\ee
where $\epsilon^{\prime} = \epsilon - \bar{g}$. We find that,
up to a shift in energy by $\bar{g}$, our result is identical to
$\rho_0 (E)$ near its band center (c.f.  Eq.(48) in Ref.  \cite{BGI}).
Therefore, in the large-$\bar{g}$ limit, the DOS of each spin
component in our spin-$\frac{1}{2}$ system seems to behave as the
shifted $\rho_0 (E)$.  However, we will show below that this is not
the case.

Now let us turn to the region at the band center.
For $\ep=0$, i.e., $x_{\pm}=\pm\bar{g}/\Ga$, we have
$f_+=f_-=e^{-2(\bar{g}/\Ga)^2},$ and
$F_+=-F_-=e^{-2(\bar{g}/\Ga)^2}
\int_0^{\sqrt{2}\bar{g}/\Ga}dxe^{x^2}.$
By Eq.(\ref{asympt}), it follows that
\be
\rho_{\uparrow}(E)|_{\epsilon =0}
\cong
\frac{\ka^2}{\pi^{5/2}\Ga_0} \frac{\pi}{2} (\frac{\bar{g}}{\Ga_0})^2
e^{-(\bar{g}/\Ga_0)^2},
\ee
which is only one-half of the value of
$\rho_0 (E) |_{\epsilon = \pm \bar{g}}$ for $\bar{g}/\Gamma_0 \gg 1$
(c.f. Eq.(47) in Ref. \cite{BGI}). It indicates that the behavior of
$\rho_\uparrow (E)$ ( and $\rho_\downarrow (E)$, due to the mirror
symmetry ) far from the peak is {\it not} identical to that of
$\rho_0  (E)$ in the band tail.

We can show the difference in the asymptotic behavior between $\rho_0
(E)$ and $\rho_\uparrow (E)$ ( $\rho_\downarrow (E)$ ) by considering
the regions around the band tails directly. For
$(\ep - \bar{g})/\Gamma \ge \bar{g}/\Gamma \gg 1$, namely,
$x_+ \gg x_- \gg 1$, from Eqs.(\ref{upay}) and (\ref{asympt}), we
obtain
\be
\rho_{\uparrow}(E)
\cong
\frac{\ka^2}{\pi^{5/2}\Ga_0} \frac{\pi}{2}
(\frac{\ep^\prime}{\Ga_0})^2
(\frac{\ep^\prime + 2\bar{g}}{\bar{g}})
e^{-(\ep^\prime/\Ga_0)^2},
\qquad \epsilon^\prime \to \infty.
\label{ep1}
\ee
Due to the mirror symmetry about $\epsilon =0$, the behavior of
$\rho_{\uparrow}(E)$ in the limit $\ep^\prime \to -\infty$ ( or
equivalently, $\ep \to -\infty$ ) can be inferred by that of
$\rho_{\downarrow}(E)$ in the limit $\ep^\prime \to \infty$.
Therefore, by Eqs.(\ref{downay}) and (\ref{asympt}), we have
\be
\rho_{\uparrow}(E)\cong
\frac{\ka^2}{\pi^{5/2}\Ga_0} \frac{\pi}{8}
(\frac{\ep^\prime + 2\bar{g}}{\bar{g}})^2
e^{-(\ep^\prime/\Ga_0)^2},
\qquad \epsilon^\prime \to -\infty,
\label{ep2}
\ee
which diminishes faster than that in the opposite side of the band tail.
Hence, the asymptotic behavior in both limits,
$\epsilon^\prime \to  \pm \infty$, does {\it not} reduce to that of
the shifted $\rho_0 (E)$ in the band tails ( c.f.  Eq.(47) of
Ref.\cite{BGI} ), as indicated in the discussion for $\epsilon = 0$.
Moreover, we find that the asymptotic behavior of $\rho_\uparrow (E)$
( $\rho_\downarrow (E)$ ) is {\it not} symmetric with respect to the
position of its peak.  Thus $\rho_\uparrow (E)$ and
$\rho_\downarrow (E)$ do {\it not} evolve to the identical form
of the shifted $\rho_0 (E)$ in all regions of energy, although they do
so near the location of their peaks.

\section{Conclusion and Discussion}

In summary, we extend the exact solution obtained by Br\'ezin {\it et
al.} for the DOS of spinless electrons at the lowest Landau level with
a short-range disorder to the cases of a spin-$\frac{1}{2}$ system.
In this generalization, we obtain an analytical expression for DOS of
spin-$\frac{1}{2}$ system, illustrating the interplay between the
Zeeman splitting of Landau levels and the disorder-induced broadening.
We also find that the broadening and the band splitting of the Landau
bands are enhanced due to the level repulsion from the mixing of two
spin degrees of freedom by the random scatterings.

We notice that a similar expression of DOS is recently reached by
Shahbazyan and Raikh \cite{SR} in the context of the double-layer
system with a tunneling coupling constant $t$.  Alternatively, we can
compare these two systems explicitly by calculating their DOS using
the path-integral approach, rather than summing up the entire diagrams
in Ref.\cite{SR}.  After introducing the holomorphic superfields
$\Psi_i$ and their adjoints $\bar{\Psi}_i$ ( $i=1, 2$ ) for electrons
in each layer, the `` action " of the double-layer system,
$S_{\rm  D}$, becomes ( c.f.  Eq.(\ref{S}) )
\end{multicols}
\bea
S_{\rm D}&=&\frac{2\pi i}{\kappa^2}\int dzdz^*d\theta d\bar{\theta}
    e^{-\frac{1}{2}\kappa^2(|z|^2+\th\bar{\th})}
   \left[ \ep (\bar{\Psi}_1 \Psi_1 + \bar{\Psi}_2 \Psi_2 )
     -t (\bar{\Psi}_1 \Psi_2 + \bar{\Psi}_2 \Psi_1 ) \right]
                        \nonumber  \\
 && +\int dzdz^*
    g_1\left[  e^{-\frac{1}{2}\kappa^2|z|^2}\frac{2\pi}{\kappa^2}
    \int d\theta d\bar{\theta} e^{-\frac{1}{2}\kappa^2\th\bar{\th}}
      \bar{\Psi}_1 \Psi_1 \right]
\label{SD}     \\
 && + \int dzdz^*
    g_2\left[  e^{-\frac{1}{2}\kappa^2|z|^2}\frac{2\pi}{\kappa^2}
    \int d\theta d\bar{\theta} e^{-\frac{1}{2}\kappa^2\th\bar{\th}}
      \bar{\Psi}_2 \Psi_2 \right] ,
                        \nonumber
\eea
\begin{multicols}{2}
where $g_1(\alpha)$ and $g_2(\alpha)$ are the corresponding Fourier
transforms of the distribution functions for the random potentials in
each layer.  For the white-noise distributions of the same strength,
by Eq.(\ref{id}), $S_{\rm D}$ reduces to the same form as our
supersymmetric `` action " in Eq.(\ref{superS}), and then leads to a
similar expression of DOS.  Thus, by the change of variables in
Eqs.(\ref{change1}) and (\ref{change2}), our model can be related to
the double-layer one.  However, the equivalence between these two
models is {\it not} held in general.  If the random distributions are
not white-noise type of the same strength, the cancellation of the
four fermion interactions will fail, and then there is no
supersymmetry for the spin-$\frac{1}{2}$ case.  However, for the
double-layer system, a supersymmetric `` action " for $S_{\rm D}$ can
still be obtained ( because there is no four fermion interaction in
$S_{\rm D}$ ), and an exact expression of DOS can be formulated for
any zero-range random distributions by a straightward generalization
of the work of Br\'ezin {\it et al.} \cite{BGI} That is, once the
disorder is included, these two systems are {\it not} exactly the same
and, moreover, for the spin-$\frac{1}{2}$ model, it is quite limited
to have an exact expression of DOS.

\noindent
{\bf Acknowledgments:}
The authors are grateful to Dr. M.-C. Chang and Dr. S.-C. Gou for
helpful discussions.  M.-H.  Tu acknowledges support from National
Science Council by Grand No.  NSC-86-2112-M-007-020.

\end{multicols}

\begin{references}
\bibitem{note} If we set $\epsilon \to -\epsilon$, then
$\nu_\pm \to -\nu_\mp,$ and, by Eqs.(\ref{zr}) and (\ref{zi}), we have
$Z_0^{\rm R} \to  Z_0^{\rm R}, Z_0^{\rm I} \to -Z_0^{\rm I}$, and
$\frac{\partial}{\partial \ep} ( Z_0^{\rm I} / Z_0^{\rm R} ) \to
 \frac{\partial}{\partial \ep} ( Z_0^{\rm I} / Z_0^{\rm R} )$.
Thus, after rewriting Eq.(\ref{exact}) as
$\rho(E)=\frac{\kappa^2}{2\pi^2}
\frac{\partial}{\partial \ep} ( Z_0^{\rm I} / Z_0^{\rm R} ) /
\left[ 1 + ( Z_0^{\rm I} / Z_0^{\rm R} )^2 \right]$, the symmetry with
respect to $\epsilon \to -\epsilon$ becomes obvious.
\end{references}
\end{document}